\documentclass[10pt,a4paper]{aastex}

\usepackage[T1]{fontenc}
\usepackage[bitstream-charter]{mathdesign}

\usepackage[latin1]{inputenc}							
\usepackage{amsmath}									

\title{Spatio-temporal Scaling of Turbulent Photospheric Line-of-Sight Magnetic Field in Active Region NOAA 11158\
       \vspace{0.5cm}}

\author{%
	Jordan A. Guerra, Antti Pulkkinen, Vadim M. Uritsky, and Seiji Yashiro \\
	Catholic University of America and NASA Goddard Space Flight Center \\
	jordan.guerraaguilera@nasa.gov \\
	\vspace{20pt}
	}

\usepackage{graphicx}
\usepackage{natbib}
\begin{document}



\begin{abstract}
\noindent 
We study structure and dynamics of turbulent photospheric magnetic field in active region NOAA 11158 by characterizing spatial and temporal scaling properties of the line-of-sight (LOS) component. Using high-resolution high-cadence LOS magnetograms from SDO/HMI, we measured power-law exponents $\alpha$ and $\beta$ describing Fourier power spectra in wavenumber ($k$) and frequency ($f$) domains and investigated their evolution during the passage of the active region through the field of view of HMI. The flaring active region NOAA 11158 produces a one-dimensional spatial power spectral density that follows approximately a $k^{-2}$ power law -- a spectrum that suggests parallel MHD fluctuations in an anisotropic turbulent medium. In addition, we found that the values of $\alpha$ capture systematic changes in the configuration of LOS photospheric magnetic field during flaring activity in the corona. Position-dependent values of the temporal scaling exponent $\beta$ showed that, on average, the core of the active region scales with $\beta >$ 3 surrounded by a diffusive region with an approximately $f^{-2}$-type spectrum. Our results indicate that only about 1 - 3 \% of the studied LOS photospheric magnetic flux displays $\beta\approx\alpha$, implying that Taylor's hypothesis of frozen-in-flow turbulence is typically invalid for this scalar field in the presence of turbulent photospheric flows. In consequence, both spatial and temporal variations of the plasma and magnetic field must be included in a complete description of the turbulent evolution of active regions. 

\end{abstract}

%
\keywords{Active regions $\cdot$ Flares, relation to magnetic field $\cdot$ Magnetic fields, photosphere $\cdot$ Photospheric turbulence}


\section{Introduction}
Solar active regions (ARs) are the central building blocks in the path to understanding the drivers of space weather. Major solar flares and coronal mass ejections (CME) originate from active regions, where strong ($\approx 10 ^{3}$ gauss (G)) and complex magnetic field structures can accumulate sufficient free energy to power energetic eruptions. The high degree of complexity in terms of topology and spatial distribution present in the photospheric magnetic field appears to emerge from a turbulent photospheric plasma state (\cite{2005ApJ...629.1141A}, \cite{2010ApJ...720..717A}). In this state, field emergence, fragmentation, and dissipation associated with turbulent flows lead to highly irregular spatio-temporal distribution of the magnetic field. In a simple way, we can view an AR as a system that takes the magnetic field and evolves it into an unstable non-potential configuration by non-linear shear and stress. For this system to return to a lower-energy state, the excess free energy must be released in a bursty event in the corona while electrical currents are dissipated and potential field configuration is restored (\cite{2011LRSP....8....6S}).

Non-linear dynamical processes, such as turbulence, are often studied using a description that involves statistical momenta of the turbulent field. For instance, in hydrodynamics (HD) and magnetohydrodynamics (MHD), kinetic and/or magnetic energy injection, transfer, and dissipation processes in turbulent flows are understood in terms of the scale-free behavior of their Fourier spectrum (\cite{1993noma.book.....B}, \cite{2011soca.book.....A}), which follows a power-law distribution in space and time. Kolmogorov's 5/3 law is a classic example of these phenomena \cite{1941DoSSR..30..301K}. In the case of the photospheric magnetic field, statistical parametric analyses have been performed with the aim of quantifying the complexity present in the field (see \cite{2005ApJ...629.1141A}; \cite{2010AdSpR..45.1067M}). However, only recently, when better and more accurate measurements of the photospheric magnetic field have become available, a more coherent picture of its complexity has started to emerge (see \cite{2005ApJ...629.1141A}).

Previous studies on the complexity in the photospheric magnetic field in ARs can be divided into two categories: (1) analysis of physical and statistical parameters of the magnetic field such as the effective connected magnetic field strength (\cite{2007ApJ...661L.109G}, \cite{2008GeoRL..3506S02G}), the strong gradient length (\cite{2002ApJ...569.1016F}, \citeyear{2003JGRA..108.1380F}), or the statistical momenta of the field spatial distribution (\cite{2003ApJ...595.1277L}, \citeyear{2003ApJ...595.1296L}; \cite{2008ApJ...688L.107B}), and (2) description of magnetic structures based on transformations of the LOS component such as the spatial power scaling exponent (Fourier analysis; \cite{2005ApJ...629.1141A}; \cite{2010ApJ...720..717A}) or fractal dimension (wavelet analysis; \cite{2010AdSpR..45.1067M}). For example, \cite{2007ApJ...661L.109G} defined the AR effective connected magnetic field strength $B_{\rm eff}$ as a measure of magnetic field complexity. The $B_{\rm eff}$ parameter accounts for the connectivity of individual photospheric magnetic flux concentrations; therefore its value depends on the spatial distribution of the flux concentrations. Values of $B_{\rm eff}$ were measured using LOS magnetograms averaged over 12 h -- a cadence too low in order to capture transient phenomena of magnetic concentrations, which encompass a wide range of temporal scales (\cite{2012ApJ...748...60U,uritsky13}).  \cite{2005ApJ...629.1141A} analyzed a sample of ARs using photospheric magnetic data from the {\it Michelson Doppler Imager} (MDI; \cite{1995SoPh..162..129S}) instrument onboard SoHO and the Digital Magnetograph (DMG) located at the Big Bear Solar Observatory. This study was focused on measuring power-law scaling exponents of the spatial power spectral densities. \cite{2005ApJ...629.1141A} concluded that the derived exponents described the scale-free behavior of the magnetic field and served as indicators for differentiating between ARs that are prone to produce flaring activity and those that are flare-quiet.

A common approach in the studies mentioned above was to quantify the complexity present in the instantaneous spatial distribution of the photospheric field and then to observe subsequent time evolution of the spatial parameter. Consequently, spatial and temporal domain analyses have been conducted for the most part in an independent fashion. A question that then naturally arises concerns the coupling between these two domains: is there a way to link the spatial and temporal variations? The first step to study this coupling is to verify the possible validity of Taylor's hypothesis of frozen-in-flow turbulence (\cite{Taylor18021938}) for the photospheric plasma. If the hypothesis is valid, determining the (temporal) spatial scaling ({\it e.g.} \cite{2002ApJ...577..487A}) is sufficient since the (spatial) temporal scaling is constrained to be identical. On the other hand, if the hypothesis is not valid, both spatial and temporal scaling must be considered in the analysis in order to provide a complete picture of the state of the photospheric magnetic field and plasma. In this report, we will demonstrate that the latter is indeed the case for NOAA AR 11158.

In this paper, we extend previous studies of solar AR magnetic field complexity by addressing both spatial and temporal variability of the LOS photospheric magnetic field across a wide range of scales. By constructing a more comprehensive picture of the turbulent spatio-temporal dynamics in the AR photospheric magnetic field, we will provide new information that can help to better understand, for example, magnetic energy release signatures in the photosphere and the coupling between the photosphere and corona. In Section 2, we describe the analyzed set of LOS magnetograms and the active region to which they belong, NOAA 11158. Section 3 explains the data analysis and discusses the results. We explain the method of measuring the power-law exponents in two separate subsections: the spatial scaling analysis (Section 3.1) and the temporal scaling analysis (Section 3.3), both based on the Fourier transform of LOS magnetic field. Our main results are reported and discussed in Sections 3.2 and 3.4, while in Section 4 we draw conclusions and outline future work.

\section{Data}
\subsection{Active Region NOAA 11158}
The first X-class flare of solar cycle 24 was generated by NOAA AR 11158 on 15 February 2011 \cite{2011ApJ...738..167S}. This AR containing two bipolar regions emerged on 11 February 2011 and was initially classified as a $\beta-$region according to the McIntosh classification (\cite{McIntosh1990}). As the AR evolved, the two bipolar regions were seen colliding and then sliding along each other, forming a $\beta\gamma-$complex sunspot group (\cite{2011ApJ...738..167S}). 

\begin{figure}[!h]
\begin{center}
\includegraphics[width=12cm]{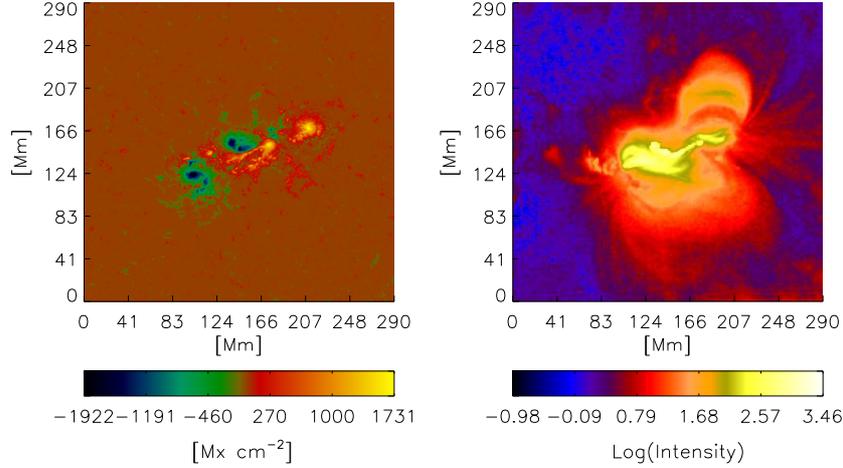}
\caption{%
Spatial distribution of SDO/HMI line-of-sight magnetic field (left) and SDO/AIA 94 {\AA} coronal emission (right) in NOAA AR 11158 during the X2.2 flare of 15 February 2011 at 01:44 UT (onset time). The active region had a quadrupole line-of-sight magnetic configuration at the time of the flare.
}
\label{AR_hmi_aia}
\end{center}
\end{figure}

At the time of the major flare, NOAA AR 11158 displayed three regions of intense magnetic field (see Figure \ref{AR_hmi_aia}): an eastern region of negative magnetic polarity, a western region of positive magnetic polarity, and a central region where a well defined {\it polarity inversion line} (PIL) separated the negative (north) from the positive (south) polarities \cite{2012AN....333..125B}. This AR has been widely studied (\cite{2011ApJ...738..167S, 2012AN....333..125B,2012ApJ...745L..17W}) because of the availability of data from many different instruments, and in particular high spatio-temporal resolution measurements from the {\it Solar Dynamics Observatory} (SDO; \cite{2012SoPh..275....3P}). In addition to the X2.2-class flare, six M-class flares of different intensities were produced by NOAA AR 11158. Table \ref{table} lists all seven flares indicating their classes, peak times, and rise times ({\it i.e.} time scale of the impulsive phase) according to the GOES X-ray flux (1 - 8 {\AA}). For each flare, the GOES X-ray flux integrated over the corresponding rise time is reported in Table \ref{table} as well.

\subsection{Data Set}
We used a set of 4780 LOS magnetograms that spans 10 days, from 10 February to February 20 2011, after which the AR disappeared from the field of view over the western solar limb. Magnetograms of 800 $\times$ 800 pixels displaying the AR were recorded with the {\it Heliospheric and Magnetic Imager} (HMI; \cite{sdo_hmi}) on board the SDO satellite at a spatial resolution of 0.5 arcsec and a temporal resolution of 3 min. 

The HMI instrument produces filtergrams measuring the Stokes parameters at different positions in the spectral line Fe {\sc i} 617.3 nm. LOS magnetograms are calculated from the Stokes parameters (\cite{StokesVector}), producing full-disk maps of 4096$\times$4096 pixels a pixel size of $0.5''$ per pixel and in a cadence of 45 s. Magnetograms were corrected by using the IDL {\ttfamily drot\_{}map} procedure in the {\it SolarSoft} package. Each HMI magnetogram was remapped to the central meridian. Assuming rigid rotation, the rotation rate was estimated to be 14.326 degree per day. At the central meridian position, the LOS component differs from the vertical component ($B_{z}$) by a factor of $\cos{\theta}$, with $\theta$ being the heliographic longitude at the center of the AR. Therefore, the LOS component was corrected by multiplying it by $[\cos\theta]^{-1}$ in order to evaluate the vertical component, assuming the field is mostly radial (see \cite{2012ApJ...761...86V} and references therein). 

Flaring activity in the AR was investigated also using the data from the {\it Atmospheric Imaging Assembly} (AIA; \cite{sdo_aia}) aboard SDO. The AIA consists of four telescopes which measure the coronal emission in 10 channels (EUV and UV) covering several emission lines of Fe ions as well as the continuum for coalignment with other data ({\it e.g.} SDO/HMI, SoHO, and TRACE). The AIA instrument enables high-resolution full-disk imaging of the corona and transition region, with a temperature range of 0.06 to 20 MK, with a cadence of 12 s, and spatial resolution of 1.5'' per pixel. For the present work, we selected the 94 {\AA} channel, corresponding to the Fe {\sc xviii} transition, which allows mapping of the flaring corona. In the Level 1.5 of the AIA data (\cite{sdo_aia}) maps are already derotated, plate-scale adjusted, and shifted to place the center of the Sun in the middle of the data array. Additionally, the coronal emission maps (Figure \ref{AR_hmi_aia}, right) were spatially and temporally coaligned with the HMI magnetograms (Figure \ref{AR_hmi_aia}, left), in order to correlate localized transient coronal features with the possible corresponding features in the photospheric magnetic field. We used GOES X-ray flux of 1 min cadence to identify the onset and the peak times of all the detected flares.

\section{Results and Discussion}
The scaling properties of magnetic flux in ARs can be studied by determining the behavior of the power spectral density (PSD) as a function of either spatial wavenumber ({\it e.g.} \cite{2010ApJ...720..717A}) or temporal frequency. Power spectral density is calculated as the square of the absolute value of the Fourier spectrum, {\it i.e.} $E(\nu)$=$|F(\nu)|^2$, where $\nu$ is either the Fourier frequency ($f$) or wavenumber ($k$). The Fourier spectrum is obtained by using the fast Fourier transform (FFT) method. Maximum and minimum measurable wavenumbers and frequencies for the spectrum are constrained by the image size, spatial resolution, temporal resolution, and total length of the studied time series. When determining the possible scale invariance in the magnetic field variations, we focus on the ranges of scale where the PSD approximately follow a power law. In Sections 3.1 and 3.3 we describe the spatial and temporal PSD analyses more in detail.

\subsection{Spatial Scaling}
The instantaneous magnetic field distribution in each LOS magnetogram (see Figure \ref{AR_hmi_aia}, left) is a function of the horizontal and vertical positions, $B_{\rm LOS}=B(x,y)$. Calculating the power spectrum of $B(x,y)$ results in a two-dimensional PSD, $E(k_{x},k_{y})$, in which
\begin{equation}
k_{x,y}=\frac{n_{i,j}}{N_{x,y}\Delta_{x,y}}=\frac{n_{i,j}}{N_{s}\Delta_{s}}; \quad \quad  n_{i,j}=0,1,...,\frac{N_{x,y}}{2},
\end{equation}
where in our case $\Delta_{x}=\Delta_{y}=\Delta_{s}$= 0.375  Mm and $N_{x}=N_{y}=N_{s}=$ 800. We defined the zero (lowest) wavenumber as $0.5/N_{s}\Delta_{s}$.

One-dimensional (1D) PSD, $E'(k)$, can be obtained from $E(k_{x},k_{y})$ by integrating the latter over the angular direction. Following \cite{2001SoPh..201..225A}, we integrated the 2D PSD over annuli defined by the circles $k$ and $k+\Delta k$ where $k=|k|=\sqrt{k_{x}^2 + k_{y}^2}$. In order to express $E'(k)$ in correct units and to be consistent with the energy constraint (see Equations (1) - (5) in \cite{2012A&A...541A..17S}), a correction of $2\pi k$ must be applied, that is, $E(k)=2\pi k E'(k)$. Here, $E(k)$ is the spectral power of the spatial magnetic field fluctuations associated with photospheric structures of the linear length $l=k^{-1}$. By applying the above transformation to each magnetogram in the data set, we construct a time-varying PSD or dynamic spectrogram, $E(k,t)$.

Scale-invariant phenomena in non-linear dynamical systems are often characterized in terms of power-law distributions (see \cite{2011soca.book.....A}, Chapter 1 for examples of systems giving rise to power laws in physical sciences). Here we determine a power law in the wavenumber domain and follow the evolution of the exponent over time. That is
\begin{equation}
E(k,t) \sim k^{-\alpha(t)},
\label{psd_k}
\end{equation}
The scaling exponent $\alpha$ is defined as the slope in a log-log representation of the data ({\it e.g.} \cite{2005ApJ...629.1141A}). The exponent $\alpha$  is measured within a range of $k$ values that correspond to the estimated {\it inertial range} of photospheric turbulence (\cite{1993noma.book.....B}). The inertial range involves scales which are smaller than the energy injection scale $l_{\rm i}$ and greater than the dissipation scale $l_{\rm d}$ (\cite{1989PhFlB...1.1964B, 1993noma.book.....B,2002ApJ...577..487A}).

\begin{figure}[!h]
\begin{center}
\includegraphics[width=10cm]{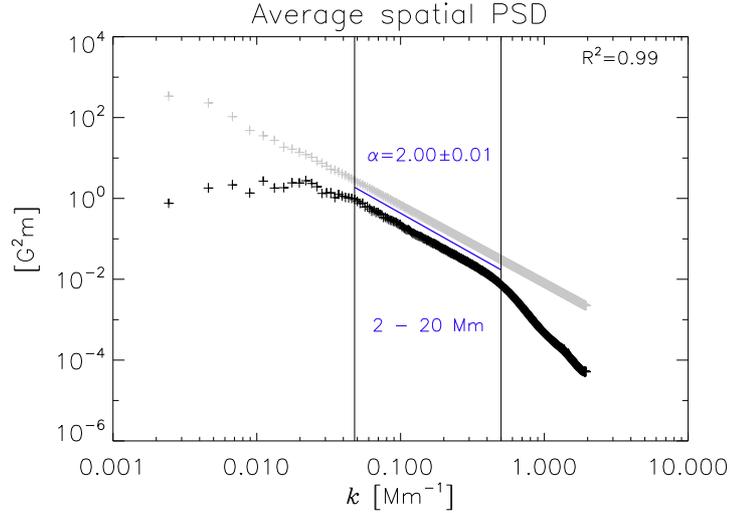}
\caption{%
Power-law behavior of the average spatial power spectral density, $\overline{E}(k)$, as a function of isotropic wavenumber (black). The power-law exponent is determined within the inertial range of wavenumbers, marked by vertical lines. The inertial range shows scaling with $\overline{\alpha}=$ 2.00 $\pm$ 0.01. The grey curve shows the spatial spectral power density of a synthetic data set with the scaling exponent of $\approx$ 1.92 $\pm$ 0.01. See text for details. Coefficient $R^{2}$ estimates the goodness of linear fit and it is displayed in the figure. The uncertainty of measure $\pm0.01$ represents the 1-$\sigma$ standard deviation.
}
\label{psd_k_ave}
\end{center}
\end{figure}

The time average of $E(k,t)$ follows a power law in the wavenumber domain as well, $\overline{E}(k)\sim k^{-\overline{\alpha}}$ in which $\overline{\alpha}$ is the scaling exponent of the time-averaged spectrum. Figure \ref{psd_k_ave} shows $\overline{E}(k)$.  The best linear fit to this log-log graph can be obtained for the wavenumber range between $k_{\rm min}$ = 0.05 Mm$^{-1}$ and $k_{\rm max}$ = 0.5 Mm$^{-1}$ (vertical solid lines in Figure \ref{psd_k_ave}), or $l\approx$ 2 - 20 Mm. This inertial range of scales is defined in the high-$k$ limit by the presence of a smooth cutoff in the PSD at $\approx$ 0.5 Mm$^{-1}$. This cutoff is likely to be caused by insufficient instrumental resolution at small scales, where the PSD must be corrected by the modulation transfer function (MTF). Therefore, in order to measure the scaling exponent for uncorrected data we must restrict the inertial range for spatial scales greater than 2 Mm (see \cite{2005ApJ...629.1141A}). At the low-$k$ limit, the value of $k_{\rm min}$ can vary. We found that this value does not affect significantly the measurement of $\overline{\alpha}$, giving us a flexibility of choosing the $k_{\rm min}$ value based on the best linear fit. In order to estimate the quality of the power-law fit, we calculate $R^{2}$, the correlation coefficient between the data points in logarithmic scales and the predicted linear model. Departures from $R^{2}= 1.0$ represent a less than ideally fitted model. We observed that the best-fit range of $k$ in the temporally-averaged spatial PSD provides the best fit for measuring $\alpha$ at every time step as well. The inertial-range time-averaged scaling exponent, $\overline{\alpha}$, takes the value 2.00 $\pm$0.01 with $R^{2}=0.99.$ Uncertainty in the scaling exponent is obtained from the 1-$\sigma$ standard deviation of the linear fit in the log-log space.

To verify robustness in our implementation of the spectral analysis and identify possible non-physical effects in the determined PSD, we tested our method on a set of synthetic images. We constructed the synthetic images in such a way that they mimic the HMI data, {\it i.e.} they display fractals that have the same inertial-range scaling, dynamical range, and spatial size as the actual data. For this purpose we have used the IDL {\ttfamily fractal\_{}synth} routine which uses the Hurst exponent $H$ (\cite{2002Hergartenbook}) as input parameter. The Hurst exponent is related to $\alpha$ as $H=(\alpha-2)/2$ for a 2D fractal whose structure is continuous through the edges (fractals form a 3D closed toroidal surface). In our computations we used $H=0.0$ which corresponds to $\alpha=\overline{\alpha}$ = 2.00. The set-averaged synthetic one-dimensional PSD is plotted in grey in Figure \ref{psd_k_ave}. Importantly, this curve shows that our implementation of the spectral analysis  accurately captures the scaling through the entire range of generated synthetic data. The measured scaling exponent for the synthetic data is 1.92 $\pm$ 0.01. Differences between synthetic and physical PSD are clear for the high- and low-wavenumber ends (Figure \ref{psd_k_ave}). The spectral power for the low-$k$ values is limited by the footpoint size ($l \approx 50 - 80$ Mm) of those coronal arcades rooted in the AR (see Figure \ref{AR_hmi_aia}) and therefore the averaged PSD deviates from a straight line and becomes flat. On the other hand, as mentioned before, the insufficient spatial resolution of the HMI data causes the differences seen for high wavenumbers.

A power-law spectrum $k^{-\overline{\alpha}}$ with $\overline{\alpha}>5/3$ suggests that those arguments based on isotropic and stationary turbulence used by Kolmogorov (1941) may not be valid for the range of spatial scales considered here. When anisotropy is taken into account for MHD turbulent systems, kinetic and magnetic fluctuations display different behaviors along and perpendicular to the local mean magnetic field (\cite{1993noma.book.....B,2009ApJS..182..310S}). The observed power-law scaling, $k^{-2}$, suggests that the fluctuations producing the power spectrum $E(k)$ take place along the dominant magnetic field direction. 

\subsection{Time Evolution of $\alpha$}
\begin{figure}[!h]
\begin{center}
\includegraphics[width=12cm]{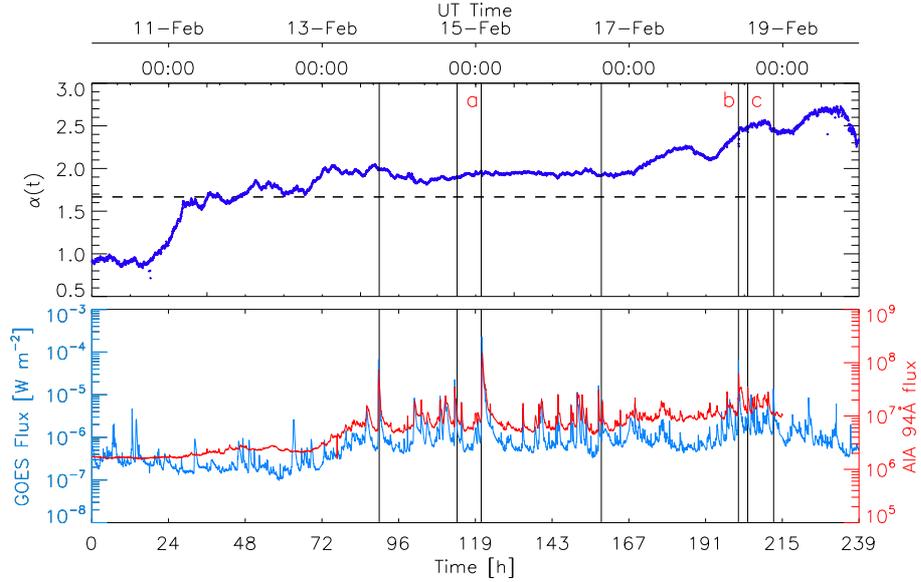}
\caption{%
Time evolution of the scaling exponent $\alpha$ (top panel) for the inertial range of spatial scales defined in Figure \ref{psd_k_ave}, and the GOES X-ray (blue) and integrated AIA 94 {\AA} (red) fluxes (bottom panel). Vertical lines indicate flare onset times. The horizontal dashed line in the top panel marks the value of the Kolmogorov exponent 5/3. The origin of the time is 9 February 2011, 23:59:19 UT.
}
\label{ab_time}
\end{center}
\end{figure}

We also tracked the evolution of the spatial scaling exponent as the AR moved across the solar disk, from the eastern to the western limb. During this time several flares were detected. Figure \ref{ab_time} displays the dynamics of inertial-range exponent $\alpha(t)$ (top panel) and the coronal emission (bottom panel), as seen by GOES X-ray flux (blue) and  AIA 94 {\AA} flux (red) integrated over the field of view. In both panels of Figure \ref{ab_time}, vertical lines mark the onset time for all seven detected M- and X-class flares.
 
Based on the evolution of $\alpha(t)$, one can associate the emergence phase of the AR with the time interval of 16--29 h beginning at 9 February 2011 23:59:19 UT. After this time, total unsigned magnetic flux continued to grow, but $\alpha(t)$ stayed within the range 1.67 to 2.0.

At the beginning of the studied time interval, the exponent $\alpha(t)$ was smaller than the Kolmogorov value of 5/3 (marked with the horizontal dashed line in Figure \ref{ab_time}). This value, $\alpha\approx 1$, might correspond to the scaling of the quiet photosphere, before the AR emerges. After about 18 h, the exponent value systematically increased until it reached a value close to 5/3. It remained close to this value for approximately 43 h and then increased to $\alpha\approx$  2. It can be argued that $\alpha\approx$ 2 is the scaling of a stable AR (\cite{2005ApJ...629.1141A}). Between 18 and 43 h, when $\alpha\approx$ 5/3, the AR could had also been in a stable state but the off-center projection effects in $B_{\rm LOS}$ could had led to lower values of $\alpha(t)$. Fluctuations of $\alpha(t)$ around $\approx$ 1.67 and 2.0 are likely due to a rearrangement of the existing photospheric flux as well as to the new emerging magnetic flux. These slow variations in $\alpha(t)$ could reach up to 10\% of the mean value, and did not exhibit clear correlation with the flaring times. In agreement with \cite{2005ApJ...629.1141A}, NOAA AR 11158, which produced a X2.2-class flare, shows an inertial-range exponent of $\alpha\approx$ 2 before this event.

\begin{figure}[!h]
\begin{center}
\includegraphics[width=11cm]{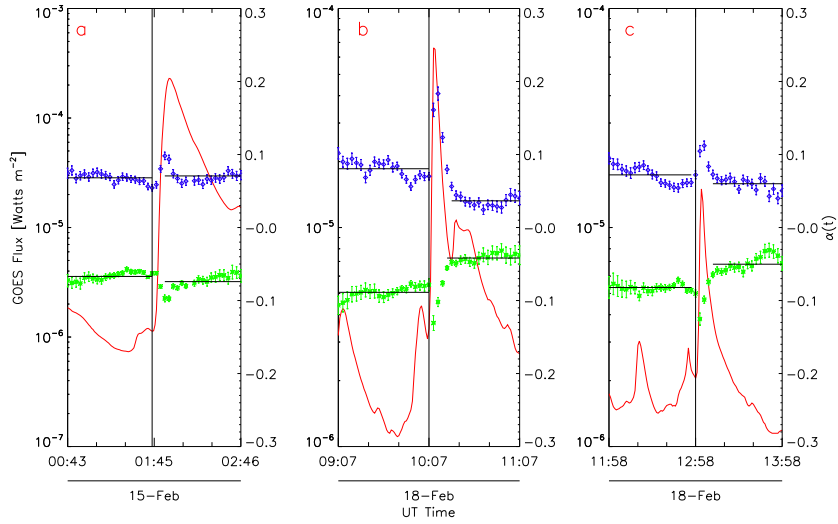}
\caption{%
Short term temporal evolution of spatial scaling exponent $\alpha$ around three flares: the X2.2-class flare (panel a) on 15 February, and two M-class flares on 18 February (panels b and c, respectively). The exponents were measured for the normalized PSD and shifted vertically ($\alpha(t)$+0.07 for blue lines, $\alpha(t)$-0.07 for green lines) for better visualization. The size of the time window of each panel is 40$\times \Delta t $(3 min)= 2 h. Panels a, b, and c correspond to labels a, b, and c in Figure \ref{ab_time}.
}
\label{zoom_flares}
\end{center}
\end{figure}

In contrast to the long-term evolution discussed above, the short term evolution of this parameter over a course of several minutes does correlate with the coronal emission time series. Figure \ref{zoom_flares} shows $\alpha(t)$ for two ranges of scales of the normalized PSD $E(k,t)/\overline{E}(k)$, where the averaged spectra were computed over the 2-h window shown in the figure. Measuring $\alpha(t)$ in the normalized spectra allows us to obtain values of the parameter that better captures changes in the spectral structure of the photospheric magnetic field. In addition to the inertial range (blue), we show $\alpha(t)$ at small $k$-values which could be associated with the energy injection range of the photospheric turbulence (\cite{1993noma.book.....B,2010ApJ...720..717A}). Each panel in Figure \ref{zoom_flares} corresponds to a flare for which we detected association between $\alpha(t)$ and GOES 1-8 {\AA}/AIA 94 {\AA} fluxes. The left panel corresponds to the X2.2-class flare observed on 15 February at 01:45 UT. The middle and right panels show a pair of M-class flares that occurred on 18 February; an M6.6-class flare with peak time 10:11 UT and an M1.5-class flare which peaked at 13:03 UT. In all three cases we observed a systematic change in both time series of $\alpha(t)$ around flaring times. From Figure \ref{zoom_flares} it can be observed that these changes are present in the magnetic field in two forms: transient or permanent \cite{2012ApJ...761...86V}. Transient changes are associated with the contamination of magnetic data (artifacts; \cite{2012ApJ...761...86V}), which are predominant during the flare impulsive phase and they display time scales comparable to those of the phase itself (see Table 1). Persistent changes, which are not related to artifacts, are evident by taking the average value of $\alpha(t)$ over 1 h, before and after the flare. We found that this average value changes after the eruption suggesting the state of the post-flare photospheric magnetic field to be different from the state of the pre-flare photospheric magnetic field. This type of permanent changes are also evident in the time series of net signed magnetic flux density when the X2.2-class flare occurred (Figure \ref{tseries} around $t=$ 119 h). For this case, a rapid increase of the total flux density was observed.

Systematic changes in $\alpha(t)$ around flaring times can be interpreted as a photospheric response to the reconnection taking place in the corona. This scenario suggests the possibility of a back-reaction from the corona to the photosphere soon after a flare takes place. The idea that the photosphere might respond to magnetic reconnection in the corona has been previously proposed and discussed. Several authors have reported observations of correlated changes in the photospheric magnetic field, both in the full vector field and LOS component (see {\it e.g.} \cite{2005ApJ...622..722L}; \cite{2010ApJ...724.1218P};  \cite{2011ApJ...727L..19L}; \cite{2012ApJ...745L..17W}; \cite{2012ApJ...761...86V}). Furthermore, we showed that spectral scaling parameters, such as $\alpha(t)$, calculated from photospheric magnetic data with sufficient spatio-temporal resolution can be used to study the photosphere-corona coupling.

\begin{table}
\caption{%
Integrated GOES X-rays flux, $F_{\rm GOES}$, and rise time $\Delta t$ for all seven detected flares of classes M and X emitted from NOAA AR 11158. The rise time corresponds to the difference between the peak and the onset time, according to GOES 1-8 {\AA} flux time series. The $F_{\rm GOES}$ flux is obtained by integrating the GOES 1 - 8 {\AA} flux over $\Delta t$ for each flare.
}
\begin{tabular}{cccc}
\hline
 GOES class & Peak time [UT] & $F_{\rm GOES}$ [J m$^{-2}$] & $\Delta t$ [min] \\
\hline\hline
M6.6 & 13 Feb., 17:38 & 1.53$\times 10^{-2}$ & 7 \\
M2.2 & 14 Feb., 17:26 & 2.27$\times 10^{-3}$ & 3 \\
X2.2 & 15 Feb., 01:56 & 5.19$\times$10$^{-2}$ & 10 \\
M1.1 & 16 Feb., 14:25 & 1.47$\times 10^{-3}$ & 4 \\
M6.6 & 18 Feb., 10:11  & 3.49$\times10^{-3}$ & 3 \\
M1.5 & 18 Feb., 13:03 &  1.29$\times10^{-3}$ & 3  \\
M1.3 & 18 Feb., 21:04 & 3.22$\times 10^{-3}$ & 9 \\
\hline
\end{tabular}
\label{table}
\end{table}

\subsection{Temporal Scaling}
Our temporal scaling analysis focuses on time series of signed magnetic flux density. First, we illustrate the method by analyzing time series of the net magnetic flux density. Second, this method is applied to a set of subfields, with sizes comparable to the typical size of a granulation cell (linear size $\approx$ 1 Mm).

\begin{figure}[!h]
\begin{center}
\includegraphics[width=12cm]{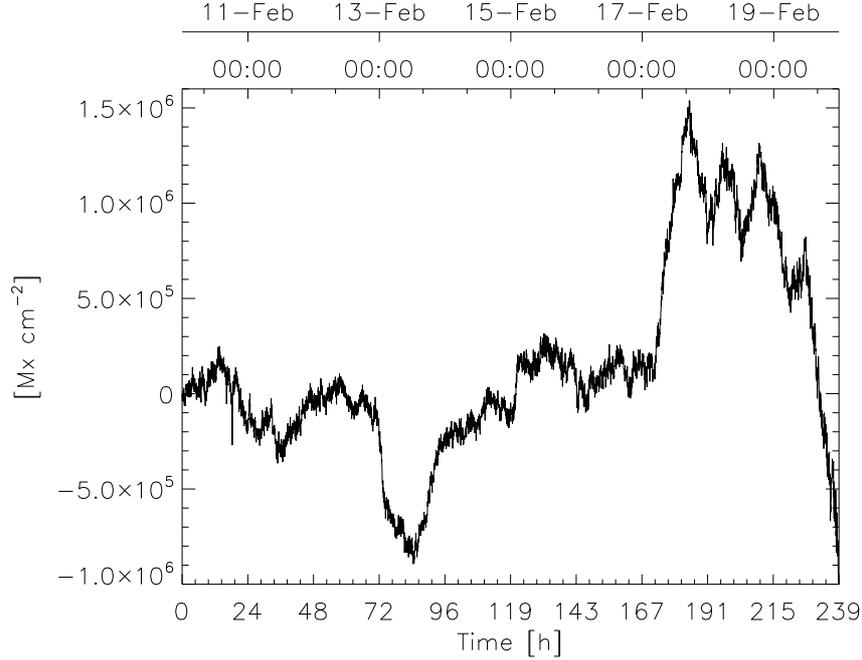}
\caption{%
Time series of the net signed magnetic flux density ($B(t)$) in NOAA AR 11158. The apparent imbalance of $B(t)$ could be caused by the limited field of view, the projection effects, or the asymmetric fragmentation leading to an underresolved magnetic flux of one polarity.
}
\label{tseries}
\end{center}
\end{figure}

We constructed the time series of the net signed magnetic flux density (Figure \ref{tseries}) by summing, at each time $t$, the contributions from all magnetogram pixels, $B(t)=\sum_{i,j} B_{\rm LOS}(x_{i},y_{j},t)$. The obtained time series was analyzed by using the FFT algorithm with a Hanning-type window \cite{1992nrfa.book.....P}. Similarly to the spatial analysis, the frequencies are defined as
\begin{equation}
f_{i}=\frac{i}{w\Delta_{t}}; \quad \quad i=0,1,...,\frac{w}{2},
\end{equation}
where $\Delta_{t}=$ 3 min is the time resolution and $w$ is the window length. The maximum measurable frequency corresponds to $f_{\rm max}=\frac{1}{2\Delta_{t}}$ while the smallest ($f_{0}$) frequency is defined as $f_{\rm min}=f_{\rm max}/w$. We applied this procedure to each point of $B(t)$ in order to construct the dynamic spectrogram, $E(f,t)$, the temporal counterpart of $E(k,t)$ defined in Section 3.1.

\begin{figure}[!h]
\begin{center}
\includegraphics[width=11cm]{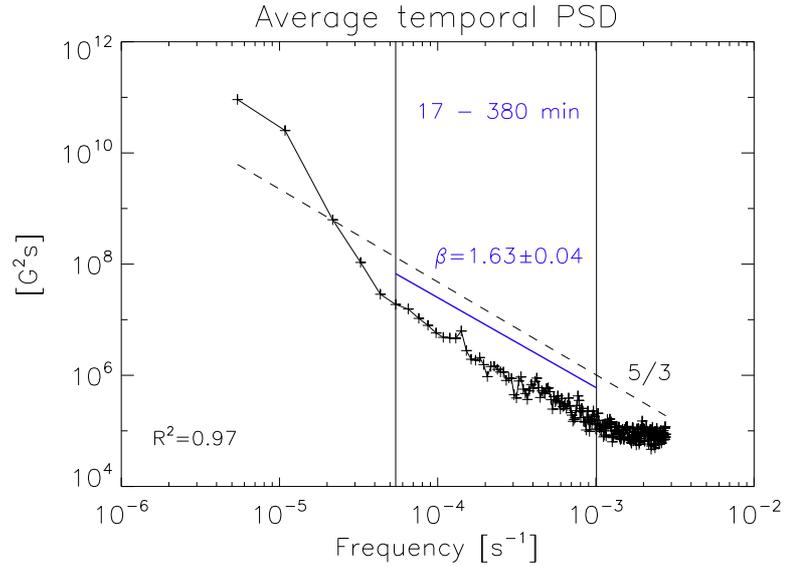}
\caption{%
Double-logarithmic plot of the averaged temporal power spectral density. Vertical lines mark the range for which we carried out the power-law fit and measured the exponent $\beta$. Photospheric LOS magnetic structures with lifetimes (or periods) between 17 and 380 min present an approximate Kolmogorov scaling in frequency ($f^{-5/3}$, black dashed curve).
}
\label{psd_f_ave}
\end{center}
\end{figure}

\begin{figure}[!h]
\begin{center}
\includegraphics[width=12cm]{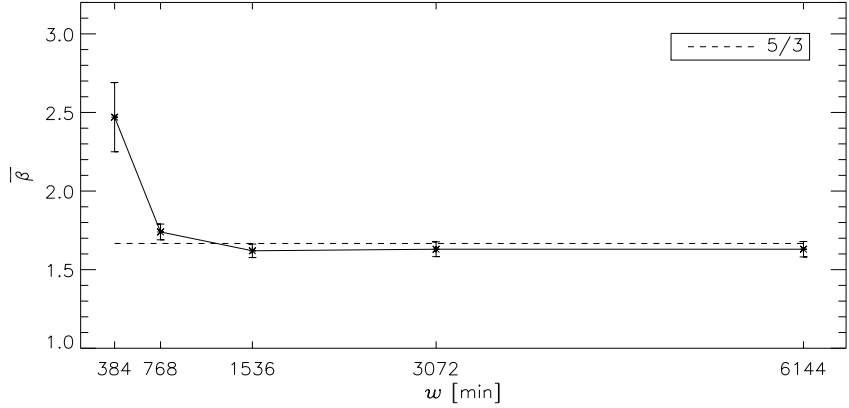}
\caption{%
Variations in the temporal scaling exponent $\overline{\beta}$ with the FFT window size $w$. Values of $\overline{\beta}$ seem to approach the Kolmogorov exponent (horizontal black dashed curve) for $w=$ 1536 min. Uncertainties are given in terms of the 1-$\sigma$ standard deviations.
}
\label{beta_vs_w}
\end{center}
\end{figure}

The power-law decay of Fourier power density was analyzed using the fit
\begin{equation}
E(f,t)\propto f^{-\beta(t)},
\end{equation}
where $\beta$ is the temporal scaling exponent, which evolves in time. This exponent characterizes non-stationary temporal autocorrelations in time series. When determining the exponent we used a window of 512 data points centered at each time point, or $w=$1536 min. This procedure was used to verify the applicability of Taylor's hypothesis in different regions (Figure \ref{beta_map_tseries}). Figure \ref{psd_f_ave} displays time-averaged temporal PSD as a function of frequency, $\overline{E}(f)$. We measured the time-averaged scaling exponent $\overline{\beta}$ of this spectrum following the same procedure as the one described in Section 3.1. The range of frequencies which provided the best fit ($R^{2}=0.97$) is 4.3$\times10^{-5}$ - 1.0$\times10^{-3}$ s$^{-1}$ or 17 to 380 min. Within this range of frequencies, $\overline{\beta}$=1.63$\pm$0.04 reveals that the time evolution of total signed magnetic flux density may follow approximately a Kolmogorov spectrum, $f^{-5/3}$ \cite{1941DoSSR..30..301K}. Several values of $w$ were tested in order to measure the temporal scaling exponent for the same frequency range (Figure \ref{beta_vs_w}). It can be seen that for $w\approx$ 1536 min, $\overline{\beta}$ already converges to $\approx$ 5/3. Therefore, longer values of $w$ do not produce significant changes in the measured exponents.

Statistical characteristics of stochastic time series are conveniently described in terms of the specific value of $\beta$, the power-law scaling exponent. For instance, $\beta=0$ corresponds to a white noise process while $\beta=2$ to a classical Brownian motion. Any other value corresponds to a fractional Brownian motion (fBm) (\cite{1982fgn..book.....M}\cite{2002Hergartenbook}). A power-law spectral density scaling with $\beta<2$ is an indication of anti-persistence in $B(t)$ following fBm, {\it i.e.} the time series reverts its tendency more often than a classical random walk, since the increments are negatively correlated. In other words, if $B(t)$ was increasing in a previous period, it is more likely that it will decrease in the next period and vice versa. This observed anti-persistent behavior may represent a balance between the supplied and dissipated photospheric magnetic fluxes involving a negative feedback. A negative feedback takes place when the output of a non-linear system is used to ``control" the input in such a way that stability can be achieved and fluctuations reduced. Based on the measured value of $\beta$, we can speculate that the photospheric plasma behaves locally as a self-regulating system: the rate at which the flux is injected into the photosphere (input) is influenced by the rate at which the flux is dissipated (output) in order to maintain a steady level of total magnetic flux. Balance between the injected and the dissipated energy is a characteristic of stationary turbulence (\cite{1993noma.book.....B}).

In addition, we believe that the negative feedback inferred from the scaling of the power spectrum is a manifestation of steady turbulence necessary for maintaining the steady state of the solar global magnetic network (\cite{2001ApJ...561..427S}). ARs are known to be embedded in the global photospheric network of magnetic elements. It has been reported that this network exists in a statically steady state with a total unsigned flux of $2-3\times 10^{23}$ maxwell (Mx) over the whole solar surface \cite{2001ApJ...561..427S}. Such a steady state can be maintained if magnetic flux is supplied at a rate of approximately $7\times 10^{22}$ Mx day$^{-1}$. According to \cite{2001ApJ...561..427S}, flux elements with intensities greater than $10^{18}$ Mx can be convected by supergranular flows towards the edges of the supergranules; there elements eventually meet with an opposite sign element and cancel out each other (annihilation). Therefore, in order to maintain the total flux intensity constant, flux elements must be injected into the photosphere at a rate that is comparable to lifetimes of the already existing flux elements. The spectrum in Figure \ref{psd_f_ave} suggests that this process is controlled by fully-developed fluid turbulence.

\subsection{Validity of Taylor's Hypothesis for Photospheric Turbulence}
It is often a challenge to measure both spatial and temporal fluctuations in turbulent flows with sufficient level of accuracy. For instance, the usage of multiple probes in plasma experiments can introduce a distortion in the flows that are observed. Less intrusive measurement techniques, such as a laser diagnostic, can provide an adequate spatial resolution, but tend to have limited temporal resolution. To address this issue, \cite{Taylor18021938} proposed a way to relate spatial and temporal characteristics of turbulence. If we study the dynamics of turbulence in terms of the Fourier representation, the temporal fluctuations $f$ on a fixed point in space associated to the passing of a Fourier mode with frequency $\omega$ and isotropic wavevector $k$ is (\cite{Moin09})
\begin{equation}
2\pi f =  \omega + {\bf v}\cdot(2\pi {\bf k}),
\label{eq5}
\end{equation}
where {\bf v} is the bulk velocity and {\bf k} is in units of m$^{-1}$. We can represent the fluid velocity field as {\bf v}={\bf U}+{\bf u'}, where the {\bf U} is the average flow speed and {\bf u}' represents the fluctuations due to turbulence. Taylor's hypothesis states that, in the presence of isotropic turbulence for which $u' \ll U$ (where $u' = |{\bf u'}|$ and $U = |{\bf U}|$), $\omega \approx$ 0. The intrinsic temporal variation of the flow can be ignored. Thus, Equation (\ref{eq5}) becomes
\begin{equation}
f \approx Uk.
\label{eq6}
\end{equation}
Therefore the temporal response at a fixed point expresses the mode $k$ that is convected through the point at the average speed of the flow, $U$. As seen from Equation (\ref{eq6}), in this way it is possible to infer the temporal behavior of turbulence from its spatial behavior and vice versa. In terms of power-law scaling, if $E(k)\sim k^{-\alpha}$ is the scaling in the spatial domain, we expect an identical power law in the frequency domain, that is, using Equation (\ref{eq6}):  $E(k=f/U)\sim (f/U)^{-\alpha}$ or $E(f)\sim f^{-\alpha}$. In other words, we expect $\beta=\alpha$ if Taylor's hypothesis holds.

In Section 3.1, we found that the time-averaged 1D spatial PSD for $B_{\rm LOS}$ approximately follows the power law $\overline{E}(k)\propto k^{-2}$ for an inertial range of scales of 0.05 - 0.50 Mm$^{-1}$. In Section 3.3, we measured the average temporal scaling for the time series of integrated flux density which follows approximately the power law $\overline{E}(f)\propto f^{-5/3}$. Because of the summation procedure used to obtain $B(t)$, the latter scaling law does not contain any spatial information. Since the two power-law exponents do not match each other, it is clear that Taylor's assumption does not hold on average. However, this assumption can still be valid at particular locations and/or during specific time intervals. This section aims at testing this possibility.  To fulfill this goal, we measured position-dependent values of the temporal scaling exponent in order to construct a $\beta$-map. Using this map, we have searched for regions where $\alpha\approx\beta$ and therefore where Taylor's hypothesis of frozen-in-flow turbulence is approximately valid.

In order to construct the map of $\beta$-values, we divided the studied field of view into non-overlapping subregions of 8 $\times$ 8 pixels, and averaged the signed flux density over these subregions for each time step. The resulting time series were analyzed as described in Section 3.3, using FFT with $w$=1536 min and a sliding window step of 150 min.

In the solar photosphere, there are several relevant wave speeds, {\it e.g.} the sound speed ($c_{\rm s}=(\gamma p_{0}/\rho_{0})^{0.5}$), the Alfv\'en speed ($v_{\rm A}=B_{0}^{2}/\sqrt{\mu\rho_{0}}$), and the slow magnetoacoustic speed ($c_{\rm t}=c_{\rm s}v_{\rm A}/(c_{\rm s}^{2}+v_{\rm A}^{2})^{0.5}$) (\cite{2003eaa..book.....M}). Each of these speeds represents a different physical mechanism of transferring energy and information. In the above definitions, $\rho_{0}$, $p_{0}$, and $B_{0}$ are characteristic values of plasma density, pressure, and magnetic field strength in the photosphere, while $\gamma$ and $\mu$ are the ratio of specific heat of the plasma and the magnetic permeability, correspondingly. For typical photospheric conditions $c_{\rm s}=$11 km s$^{-1}$, $v_{\rm A}=$12 km s$^{-1}$, and $c_{\rm t}=$ 8.1 km s$^{-1}$ (\cite{2003eaa..book.....M}). Another relevant energy transfer process is plasma convection in the photospheric plane. The horizontal component of the convection velocity, $v_{\rm hc}$, lies in the range of 0.45 - 0.50 km s$^{-1}$ (\cite{2000SoPh..193..313S}). In order to obtain the best power-law fit, we have chosen the upper limit of this range (0.50 km s$^{-1}$) as a proxy to $v_{\rm hc}$. Also, since values of $c_{\rm s}$, $v_{\rm A}$, and $c_{\rm t}$ are comparable, we have represented them all by a single value $v_{\rm MHD}$ corresponding to the average of the three speeds, 10.4 km s$^{-1}$. Substituting the values of $v_{hc}$ and $v_{MHD}$ for $U$ in Equation (\ref{eq6}), and using the inertial range $k=0.05 - 0.5$ Mm$^{-1}$ we calculated the corresponding ranges of frequencies for measuring the scaling exponent $\beta$:
\begin{itemize}
\item Range 1:  4.05$\times 10^{-4}$ - 1.5$\times 10^{-3}$ s$^{-1}$ (periods: 4 - 41 min) for $v_{\rm MHD}$,
\item Range 2:  2.50$\times 10^{-5}$ - 2.50$\times 10^{-4}$ s$^{-1}$ (periods: 66 - 660 min) for $v_{\rm hc}$.
\end{itemize}

By using these two ranges, we have divided the temporal PSD into two parts, each one described by a different scaling exponent. The low-frequency part corresponds to the mapping of the inertial range using the convection speed while the high-frequency part encompasses the MHD wave speed range. In addition to these ranges, we also measured the exponent $\beta$ over an empirical range providing the best power-law fit to the temporal PSD.

\begin{figure}[!h]
\begin{center}
\includegraphics[width=10cm]{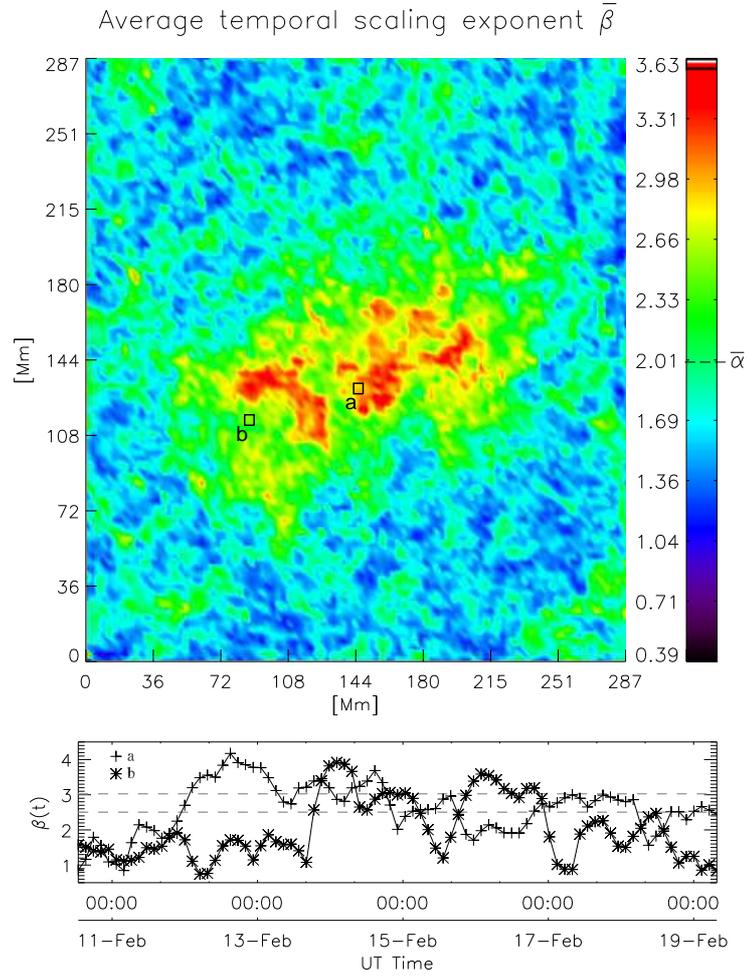}
\caption{%
Top: Spatial distribution of averaged temporal scaling exponent, $\overline{\beta}$ measured in the range of frequencies 2.5$\times10^{-5} - 2.5\times10^{-4}$ s$^{-1}$. This range  corresponds to inertial range of $k$ scales ($0.05 - 0.5 $ Mm$^{-1}$) mapped by using the nominal horizontal convection speed of the photospheric flow of 500 m $s^{-1}$. Bottom: Time series of $\beta$ on the sub-regions a and b indicated in the top panel.
}
\label{beta_map_tseries}
\end{center}
\end{figure}

Figure \ref{beta_map_tseries} displays the average map of temporal scaling exponents ($\overline{\beta}$) measured for the time-averaged temporal PSD (top panel), and time series $\beta(t)$ of instantaneous exponent values for sub-regions a and b (bottom panel). These values of $\overline{\beta}$ and $\beta(t)$ were calculated using range 2, corresponding to the horizontal convection speed. It can be seen that different areas in the field of view have different values of $\overline{\beta}$: (1) Outside of the AR, where magnetic fluctuations represent a combination of two types of stochastic behavior: flicker-type noise ($f^{-1}$, blue) and a Brownian noise ($f^{-2}$, green). At the boundary of the AR we found $\overline{\beta}\approx 2.0 - 2.3$, while in the core of the AR, $\overline{\beta}>3$. The scaling exponent of the time-averaged spatial PSD (Section 3.1), $\overline{\alpha}$ = 2.00, is shown on the color bar of Figure 7 for comparison. Photospheric regions where the scaling exponents $\overline{\alpha}$ and $\overline{\beta}$ are similar are those where Taylor's hypothesis is approximately valid. Note that time-dependent values $\beta(t)$ at locations {\it a} and {\it b} can deviate considerably from their average values $\overline{\beta}\approx 3$ and $\overline{\beta}\approx 2.5$, respectively.

In Figure \ref{beta_map_tseries}, the area of the photosphere characterized by $\overline{\beta} \approx \overline{\alpha}=$ 2.00$\pm$0.01 (see Figure \ref{psd_k_ave}) is only about 1\% of the field of view. This indicates that typically spatial and temporal scaling behavior of photospheric plasma cannot be linked by simply using Taylor's approximation. In the case of underestimation of uncertainties, we can assume them to be up to 10\% of the measured exponents. In such a case, the total area for which $\overline{\beta} \approx \overline{\alpha}$ is only about 12\%. Therefore, in order to obtain a full description of this system, independent information about spatial and temporal aspects of its behavior needs to be considered.

Similar maps of $\overline{\beta}$-values were obtained for range 1, calculated using the MHD wave speed as well as the empirical range yielding the best power-law fit. The $\overline{\beta}$ values over these ranges varied between 0.39 and 2.35. Only 3 \% of the field of view showing $\overline{\beta} \approx \overline{\alpha}$ was identified for these ranges.

\section{Conclusions}
We have investigated the turbulent state of the line-of-sight photospheric magnetic field by characterizing its Fourier spectral density, both in spatial and temporal domains. By measuring power-law spectral exponents $\alpha$ (spatial) and $\beta$ (temporal), we studied the photospheric plasma dynamics and its possible implications for the photosphere-corona coupling. In this investigation we used high spatial resolution, high cadence LOS magnetograms and maps of coronal emission from SDO/HMI and SDO/AIA, correspondingly. The utilized data represent NOAA AR 11158.

We determined the spatio-temporal scaling in two stages. First, the scaling was studied by measuring $\alpha$ for the spatial power spectral density. In the second stage we carried out the temporal analysis, in which we determined temporal scaling of time series corresponding to the net LOS magnetic field. In both stages, average and time-dependent values were measured.

Time-averaged and time-dependent values of the spatial scaling exponent were measured for the turbulence inertial range of scales, which was determined here to be the scales with linear sizes $l=k^{-1}\approx$ 2 - 20 Mm. Average power spectral density displayed a power law $\overline{E}(k)\sim k^{-\overline{\alpha}}$ with scaling exponent $\overline{\alpha}\approx 2$. In addition, the time evolution of the power-law exponent shows values greater than 5/3 during the stable phase of the AR, in agreement with \cite{2010ApJ...720..717A} for flaring ARs. On the other hand, the power law $k^{-2}$ seems to be a characteristic spectrum for MHD turbulence in which the presence of dynamically uniform strong magnetic field favors kinetic and magnetic fluctuations along the field.

Temporal spectral analysis of the data showed that the time series of net signed LOS magnetic flux density displays a power-law spectrum which can be approximated by the Kolmogorov exponent $E(f)\propto f^{-5/3}$ for an inertial range of temporal scales from several minutes to several days. Time series presenting a power-law spectra with $\beta\neq$ 2 are described in terms of the fBm model. In particular, for a fBm time series, $\beta\approx 5/3$ implies an anti-persistent behavior with weakly-anticorrelated increments. In the context of photospheric magnetic field evolution this fBm behavior could be an indication of the system seeking for a balance between injection and dissipation of the photospheric magnetic flux. We believe this balance is a signature of fully-developed turbulence that controls the photospheric magnetic flux dynamics and it is present in order to maintain the statistically steady state of the global photospheric magnetic network (\cite{2001ApJ...561..427S}).

Position-dependent average values of temporal scaling exponent ($\overline{\beta}$) indicate that regions with high average magnetic flux densities are typically associated with higher spectral slopes. Exponents $\overline{\beta}$ were measured for a range of frequencies corresponding to the spatial inertial range and consistent with Taylor's approximation. We found that only 1 - 3 \% of the studied image area satisfies the condition $\overline{\beta}\approx\overline{\alpha}$. This implies that Taylor's frozen-in-flow turbulence hypothesis is invalid for most of the field of view, including the AR. Consequently, a linear mapping between spatial and temporal behavior using Taylor's hypothesis seems questionable, and a full spatio-temporal characterization of the photospheric magnetic field is required for a complete description of the system turbulent dynamics. We have taken the first initial steps towards such spatio-temporal characterization in this work.

Short-term evolution (minutes to a few hours) of spatial scaling exponent $\alpha(t)$ captures systematic changes in the spatial distribution of LOS photospheric magnetic field associated with flaring activity. Flare-related changes manifest themselves in both transient (9 - 12 min) and persistent ($\approx$ 1 h or longer) variations of $\alpha(t)$ at the time of the flare and immediately afterwards, respectively. Transient variations in $\alpha(t)$ are most likely associated with artifacts in the magnetic field data, while persistent changes suggest a change in the state of the photospheric field. NOAA AR 11158 produced six M-class flares during its passage through the field of view of the instrument. We detected such systematic changes for two of these flares, in addition to the X-class flare. Although our results support the idea of a back reaction from the corona to the photosphere right after a flare occurs, careful analysis of a larger data set is required to confirm what types of flares are capable of influencing the post-flare state of the photospheric magnetic field.

Careful spatio-temporal analysis of high-resolution photospheric and coronal images such as the one conducted in this study can improve our understating of the physics of solar ARs and the links between the photosphere and corona during flaring activity. Furthermore, photospheric parameters such as the scaling exponents $\alpha$ and $\beta$ may also contain advanced information about the coronal flaring activity. We will expand our studies initiated in this paper by inclusion of new ARs into the analysis and by using full vector photospheric magnetic field data. Our ultimate goal is to better understand the physical properties of flaring ARs and to seek for new precursors for pending major solar eruptions.
  
%
\begin{acknowledgements}
We thank SDO/HMI and SDO/AIA teams for the data used in this study. This work was done under CEPHEUS cooperative agreement between The Catholic University of America and NASA Goddard Space Flight Center. We thank Dr Karin Muglach for useful discussions. 
\end{acknowledgements}

\bibliography{adssample}

\end{document}